\documentclass[twocolumn,prb,showpacs,preprintnumbers,amsmath,amssymb]{revtex4}
\usepackage{graphicx}
\usepackage{dcolumn}
\usepackage{bm}
\usepackage{bm}
\usepackage{latexsym}
\usepackage{amsmath}
\usepackage{amsthm}
\usepackage{amssymb}
\usepackage{amsfonts}

\newcommand{\be}{\begin{equation}}
\newcommand{\ee}{\end{equation}}
\newcommand{\bea}{\begin{eqnarray}}
\newcommand{\eea}{\end{eqnarray}}

\def\a{\alpha}
\def\b{\beta}
\def\e{\epsilon}

\def\o{\omega}

\def\G{\Gamma}
\def\D{\Delta}
\def\O{\Omega}
\def\L{\Lambda}

\def\ra{\rightarrow}

\def\bk{{\bf k}}

\def\bJ{{\bf J}}

\def\nn{\nonumber}
\def\lb{\label}
\def\pref#1{(\ref{#1})}

\newcount\bozza \bozza=1
\ifnum\bozza=0
\newdimen\shift \shift=-2truecm
\def\lb#1{%
{\label{#1}\rlap{\kern\shift{$\scriptstyle#1$}}}}
\else\def\lb#1{\label{#1}} \fi

\begin{document}

\title{Vertex renormalization in DC conductivity 
of doped chiral graphene}
\author{E. Cappelluti$^{1,2}$ and L. Benfatto$^{3,1,2}$}

\affiliation{$^1$SMC Research Center, CNR-INFM, and ISC-CNR,
via dei Taurini 19, 00185 Roma, Italy}

\affiliation{$^2$Dipart. di Fisica, Universit\`a ``La Sapienza'',
P.le A. Moro 2, 00185 Roma, Italy}

\affiliation{$^3$Centro Studi e Ricerche ``Enrico Fermi'',
via Panisperna 89/A, I-00184,
Rome, Italy}

\date{\today}

\begin{abstract}
The remarkable transport properties of graphene follow not only from the
the Dirac-like energy dispersion, but also from the chiral nature of its
excitations, which makes unclear the limits of applicability of the
standard semiclassical Boltzmann approach. In this paper we provide a
quantum derivation of the transport scattering time in graphene in the case
of electron-phonon interaction. By using the Kubo formalism, we compute
explicitly the vertex corrections to the DC conductivity by retaining the
full chiral matrix structure of graphene. We show that at least in the
regime of large chemical potential the Boltzmann picture is justified. This
result is also robust against a small sublattice inequivalence, which
partly spoils the role of chirality and leads to a doping dependence of the
resistivity that can be relevant to recent transport experiments in
doped graphene samples.
\end{abstract}

\maketitle

\section{Introduction}

The physical properties of doped and undoped graphene are the subject of an
intense investigation in the view of possible applications of these
materials in electronic and optical devices.  A precise characterization of
its transport and optical properties is thus a compelling issue. In
addition to its relevance for technological applications, graphene poses
several unusual and interesting theoretical problems.  The low energy
properties are dominated by the Dirac-like excitations at the so-called K
and K$'$ points of the Brillouin zone, where the tight-binding electronic
dispersion of the graphene honeycomb lattice can be approximated as
$\e_{\bf k} = \pm \hbar v_{\rm F}|{\bf k}|$. Here ${\bf k}$ is the
relative momentum with respect to the K (K$'$) point.\cite{guinea_review}
An additional ingredient is also the chiral character of the bands, which
gives rise to the well-known absence of backward scattering in
transport.\cite{ando_back} Note that, although obviously related in
graphene, these two issues are formally distinct and are associated with
different phenomenologies.  A Dirac-like behavior can be induced for
instance in nodal $d$-wave superconductors, as cuprates, without any
relation to chirality.\cite{lee,hirschfeld,homes} On the other hand a
chiral structure is still present even in bilayer graphene, where the low
energy bands acquire a parabolic character.\cite{mccann}

Transport properties represent a particular delicate issue in graphene.
As in any other system, also for graphene
the full quantum treatment of transport processes would require
the explicit evaluation of the current vertex corrections
for the conductivity.\cite{holstein,mahan_book}
Indeed, only going beyond the so-called bare-bubble
approximation one can recover the distinction between the transport
relaxation time $\tau_{\rm tr}$ and the ``quasi-particle'' one $\tau_{\rm
qp}$ (relevant for instance for photoemission measurements). However, the
chiral nature of excitations in graphene increases considerably the
complexity of this approach. For this reason, transport properties are
often discussed in doped graphene (i.e. when the chemical potential is far
from the Dirac point) within the framework of the semiclassical Boltzmann
theory.\cite{nomura,ando_screen,hwangprl,adam_pnas,novikov,stauber,peres,adam_bilayer,adam_ssc}
In this context, $\tau_{\rm tr}$ differs from $\tau_{\rm qp}$ for a
weighted average of the scattering probability with the angular factor
$(1-\cos\theta)$ (where $\theta$ is the angle between incoming and
outcoming scattering electrons). As usual, this leads to the suppression of
{\em forward} scattering processes.\cite{hwang_vs,hwang_ph}

Even though a general expectation holds that Boltzmann theory should be
valid for doped graphene, the application of this approach to a chiral
system as graphene is far from being trivial.  Indeed, due to the multiband
chiral structure, the velocity operator $\hat{{\bf v}}_{\bf
k}=\hbar^{-1}d\hat{H}_{\bf k}/d{\bf k}$ in graphene does not commute with
the Hamiltonian $\hat{H}_{\bf k}=\hbar v_{\rm F} {\bf k}
\cdot\mbox{\boldmath$\sigma$}$ itself, where ${\bf k}=(k_x,k_y)$ and
$\mbox{\boldmath$\sigma$}=(\hat{\sigma}_x,\hat{\sigma}_y)$ is the vector of
the Pauli matrices.  As it has been observed by many
authors,\cite{auslender,trushin1,trushin2} this fact poses several doubts
on the applicability of conventional Boltzmann theory.  In Boltzmann theory
indeed one assumes that the equilibrium distribution function $f(\e_{\bf
k})$ in the presence of the external electric field ${\bf E}$ can be
described in terms of the one in the absence of external fields as $f_{\bf
E}({\bf k}) \approx f^0(\e_{\bf k}-e\tau_{\rm tr} {\bf v}_{\bf k}\cdot {\bf
E})$, where ${\bf v}_{\bf k}=\hbar^{-1}d\e_{\bf k}/d{\bf k}$ and $e$ is the
electron charge.\cite{ziman} An important underlying assumption here is
that the energy eigenvalue $\e_{\bf k}$ is a good quantum number as well as
the shifted quantity $\e_{\bf k}-e\tau_{\rm tr} {\bf v}_{\bf k}\cdot {\bf
E}$.  This would imply that the energy Hamiltonian operator $\hat{H}_{\bf
k}$ and the velocity operator $\hat{{\bf v}}_{\bf k}$ commute so that they
can be diagonalized {\em simultaneously}.  As discussed above, however,
this condition is not fulfilled in chiral graphene.  To overcome this
problem alternative approaches based on quantum and/or quasiclassical
kinetic equations have been employed, where distribution functions and the
corresponding density operators are defined in a chiral matrix
space.\cite{auslender,trushin1,trushin2}

A second potential limit in the applicability of Boltzmann theory concerns
the origin of the angular factor $(1-\cos\theta)$ in the expression of the
transport scattering time. Indeed, in conventional systems it originates
from the momentum dependence of the current operator
$\hat{\mbox{\boldmath$j$}}({\bf k}) =(e/\hbar)d\hat{H}_{\bf k}/d{\bf k}$,
which points along the ${\bf k}$ direction in the isotropic case,
$\hat{\mbox{\boldmath$j$}}({\bf k})\propto {\bf k}/|{\bf k}|$.  It is
precisely such directional dependence which gives rise to the angular
factor $1-{\bf v}_{\bf k}\cdot{\bf v}_{\bf k'}/|{\bf v}_{\bf k}^2| \approx
1-{\bf k}\cdot{\bf k'}/|{\bf k}^2|=1-\cos\theta$ in the transport
properties.\cite{holstein,mahan_book} Things are drastically different in
the case of graphene where we have
\begin{eqnarray*}
\hat{\mbox{\boldmath$j$}}({\bf k})
&=&
\frac{e}{\hbar}\frac{d\hat{H}_{\bf k}}{d{\bf k}}
=
ev_{\rm F}\mbox{\boldmath$\sigma$}.
\end{eqnarray*}
The important feature to be stressed is that the bare vertex current
operator is here ${\bf k}$-{\em independent}, i.e. it does not depend
on the direction of the momentum ${\bf k}=(k_x,k_y)$.  Thus the
possible relevance of
the angular factor $(1-\cos\theta)$ suppressing forward scattering in the
transport properties should be better assessed.
This is particularly important when the
scattering processes involve a momentum dependence more complex that the
short-range impurity scattering discussed in
Refs.\ \onlinecite{ando_transport,falkoprl,falkoprb}.
A typical example is provided by the
scattering of electrons by acoustic phonons, which plays a major role in
controlling the temperature dependence of the DC
conductivity.\cite{fuhrer,stormer}

In order to assess these open issues we provide in this paper an explicit
derivation of the DC conductivity in doped graphene using a fully quantum
approach for the electron-phonon scattering based on the Kubo formula.
Indeed, in non-chiral systems this procedure is known to give a quantum
derivation of the Boltzmann theory.\cite{holstein,mahan_book} In the case
of graphene this program can be fulfilled only retaining the full chiral
structure in the explicit calculation of the current vertex corrections. In
the limit when the chemical potential is larger than the quasiparticle
scattering time an analytical solution can be derived.  We show that, in
spite of the above argumentations, at least in this regime the results of
the Boltzmann theory are still valid, in the sense that the
forward-scattering suppression in the transport relaxation rate is still
operative, as resulting from the additional angular factor $(1-\cos\theta)$
in $\tau_{\rm tr}$. This result follows from the fact that the dependence
on the momentum angle of the vertex function is strictly replaced by a
corresponding angular dependence in the pseudospin space, encoded in the
Pauli matrix structure. We investigate the applicability of the Boltzmann
results also in the presence of a weak inequivalence between the two carbon
sublattices, which, close to the Dirac point, leads to a mixing of the
chiral eigenstates.\cite{mucha,rotenberg_njp07} We show that even in this
case the Boltzmann results are recovered. The present quantum approach
provides in addition a basis for the future calculation of the quantum
vertex corrections at finite frequency and in the Dirac limit, which can be
both relevant for a direct comparison with existing experimental
data.\cite{fuhrer,stormer,basov}

The structure of the paper is the following. In Sec. II we introduce the
Hilbert space we are working with and the
basic formalism for the electron-phonon interaction, while
in Sec. III we evaluate the self-energy due to the
electron-phonon interaction in graphene. In Sec. IV we implement the
calculation of the vertex corrections for the DC conductivity, and we
derive the explicit expression for the transport scattering rate. In
Sec. V we report the results in the presence of an additional weak
inequivalence between the two sublattices, which breaks the chirality
preserving the Boltzmann approach. In Sec. VI we discuss some possible
observable effects of our analysis in relation to the
doping dependence of the high-temperature linear slope
of the resistivity, and in Sec. VII we summarize our conclusions.
In the Appendices A and B we include some details on the calculation of the
Eliashberg functions and on the gapped case, respectively.

\section{The model}
\label{s:model}

In order to point out the relevance of the angular transport factor
$(1-\cos\theta)$ in chiral doped graphene, we consider in this paper the
effects of electron scattering with acoustic phonons.

Let us start by introducing 
the general electron-phonon Hamiltonian
in terms of the usual orbital spinor
$\psi_{\bf k}^\dagger=
(c_{{\bf k},{\rm A}}^\dagger, c_{{\bf k},{\rm B}}^\dagger)$,
where $c_{{\bf k},{\rm A}}^\dagger$, $c_{{\bf k},{\rm B}}^\dagger$
represent the creation operator of one electron
with momentum ${\bf k}$
on the sublattices A and B respectively.
In this basis we can write:
\begin{eqnarray*}
\hat{H}
&=&
N_s\sum_{\bf k}\psi_{\bf k}^\dagger \hat{H}_{\bf k}^0 \psi_{\bf k}
+\sum_{{\bf q},\nu}\omega_{{\bf q},\nu}a_{{\bf q},\nu}^\dagger a_{{\bf q},\nu}
\nonumber\\
&&
+N_s\sum_{{\bf k,q},\nu}
\psi_{\bf k+q}^\dagger\hat{g}_{{\bf k,k+q},\nu}\psi_{\bf k}
\left(a_{{\bf q},\nu}+a_{{\bf -q},\nu}^\dagger\right).
\end{eqnarray*}
Here $\hat{H}_{\bf k}^0$ is the non-interacting tight-binding
electron Hamiltonian,
\begin{eqnarray}
\lb{h0}
\hat{H}_{\bf k}^0
&=&
\left(
\begin{array}{cc}
0 & f^*(\bk) \\
f(\bk) & 0
\end{array}
\right),
\end{eqnarray}
where $f^*(\bk)$ is the Fourier-transform of the tight-binding model
on the honeycomb lattice, $a_{{\bf q},\nu}^\dagger$ is the
creation operator of one phonon with momentum ${\bf q}$ in the $\nu$
branch, $\omega_{{\bf q},\nu}$ is the corresponding frequency and $\hat{g}_{{\bf
k,k+q},\nu}$ is the electron-phonon matrix element which presents in
general a non trivial matrix structure.  $N_s=2$ takes into account here
the spin degeneracy which will play no role in the following.

In this paper we shall work in
the original basis of electron operators for the A and B
sublattices, as described in Eq. (\ref{h0}).
This choice results to be more convenient in the evaluation
of the transport properties from a linear response theory
based on the Kubo's formula.
As an alternative approach 
the eigenvector basis, in which the Hamiltonian
(\ref{h0}) is diagonal, was employed in 
Refs.\ \cite{auslender,trushin1,trushin2} since it 
makes easier the implementation of a quantum or quasiclassical extension of
the kinetic equations.

A compelling treatment of the electron-phonon interaction including the
full phonon spectrum for all ${\bf q}$'s is a formidable task due to the
complex matrix structure of the elements $\hat{g}_{{\bf k,k+q},\nu}$.
Fortunately, since in weakly doped graphene the electron momenta are close
to the Dirac point, the relevant exchanged phonons are mainly located
either close to the K, K$'$ edge zone or at the $\Gamma$ (${\bf q}=0$)
point. Detailed analysis of the electron-phonon effects for general
acoustic and optical modes can be found in
Refs. 
[\onlinecite{castroneto,dresselhaus,manes,park,bonini,calandra,tse,basko}].
We concentrate here on the electron-phonon scattering with the acoustic
modes with ${\bf q}\approx 0$, which are dominant in the DC transport
properties at low temperatures.  Several simplifications can be employed in
this case.  First of all, the phonon dispersion can be simply linearized,
$\omega_{{\bf q},\nu}\approx \hbar v_{\rm s} |{\bf q}|$.  In addition for
${\bf q}\rightarrow 0$ the charge modulation wavelength $\lambda=1/|{\bf
q}|$ is much larger than the interatomic distance $a$.  In this regime the
two carbon atoms of the unitary cell are essentially indistinguishable so
that the electron-phonon matrix element behaves as $\hat{g}_{\bf k,k+q}
\approx g_{\bf k,k+q}\hat{I}$ in the chiral space.  Within the same
assumption, we can also neglect intervalley scattering and discuss
interactions within a single Dirac cone.\cite{ando_transport} Finally, for
small doping, we can linearize the electron dispersion close to the Dirac
points $\hat{H}_{\bf k}^0=\hbar v_{\rm F} {\bf k}
\cdot\mbox{\boldmath$\sigma$}$. We can write thus our effective Hamiltonian
as:
\begin{eqnarray}
\hat{H}
&=&
N_sN_k\sum_{\bf k}\psi_{\bf k}^\dagger 
\hbar v_{\rm F} {\bf k}
\cdot\mbox{\boldmath$\sigma$}\psi_{\bf k}
+\sum_{\bf q} \hbar v_{\rm s} |{\bf q}|
a_{\bf q}^\dagger a_{\bf q}
\nonumber\\
&&
+N_sN_K\sum_{\bf k,q}
g_{\bf q}
\psi_{\bf k+q}^\dagger\hat{I}\psi_{\bf k}
\left(a_{\bf q}+a_{\bf -q}^\dagger\right),
\label{hamdirac}
\end{eqnarray}
where we made use of the relation $g_{\bf k,k+q}\simeq g_{\bf q}$
for ${\bf q}\rightarrow 0$.

\begin{figure}[t]
\includegraphics[angle=90,scale=1]{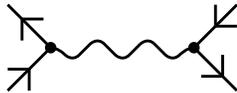}
\caption{Diagrammatic representation of the
electron-phonon interaction. Straight lines represent incoming
and outcoming electrons, the wavy line is the phonon propagator
and the filled circles are the el-ph matrix elements.}
\label{f-kernel}
\end{figure}

The resulting electron-phonon coupling
is usually expressed in terms
of the kernel $W_{\bf k-k'}(\omega-\omega')$,
which is diagrammatically depicted in Fig. \ref{f-kernel}
and which represents the effective retarded interaction
between two electrons with momenta ${\bf k}$, ${\bf k}'$
and energies $\omega$, $\omega'$, which exchange momentum
${\bf k-k'}$ and energy $\omega-\omega'$.
From (\ref{hamdirac}) we have
\begin{eqnarray}
W_{\bf k-k'}(\omega-\omega')
&=&
\left|g_{\bf k-k'}^2\right|D_{\bf k-k'}(\omega-\omega'),
\label{kw}
\end{eqnarray}
where $D_{\bf q}(\Omega)=2\omega_{\bf q}/[\omega_{\bf q}^2-\Omega]$
is the phonon propagator.
Note that, since $\hat{g}_{\bf q}\propto \hat{I}$
in the limit ${\bf q}\rightarrow 0$, Eq. (\ref{kw}) does not present
a matrix structure, simplifying notably the calculations.
It is also convenient to express the effective retarded interaction
is term of the Eliashberg function $\alpha^2F({\bf q},\Omega)$:
\begin{eqnarray}
W_{\bf k-k'}(\omega-\omega')
&=&
\int d\Omega \frac{2\Omega \alpha^2F({\bf k-k'},\Omega)}
{\Omega^2-(\omega-\omega')^2},
\label{wkernel}
\end{eqnarray}
where 
$\alpha^2F({\bf k-k'},\Omega)=|g_{\bf k-k'}|^2
\delta(\Omega-\omega_{\bf k-k'})$.

\begin{figure}[t]
\includegraphics[scale=0.3]{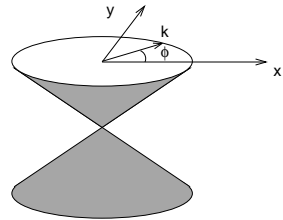}
\caption{Sketch of the cylindrical coordinates
used in the paper. In this coordinate basis we can write
${\bf k}=k(\cos\phi,\sin\phi)$.}
\label{f-axis}
\end{figure}

In this paper we shall focus on the case of intrinsically doped graphene,
where the chemical potential $|\mu|$ is much larger than both the allowed
exchanged phonon energies $\omega_{\rm max}$ and the quasiparticle
scattering rate $\Gamma_{\rm qp}$.  Note that in the case of acoustic
phonons, the highest exchanged phonon energy $\omega_{\rm max}$ is given by
the Bloch-Gr\"uneisen energy scale $\o_{\rm max}=2\hbar v_sk_{\rm F}$, so
that the constraint $|\mu| \gg \omega_{\rm max}$ implies $v_{\rm F} \gg
v_s$, which is always fulfilled in graphene.  
The condition $\G_{\rm qp} \ll |\mu|$, on the other hand,
is doping-dependent and it is usually fulfilled
in doped graphene.
In this situation, since electrons are scattered only
within a narrow energy window $\pm \omega_{\rm max}$ around the Fermi
level, we can put the electron momenta appearing in the Eliashberg function
on the Fermi surface, so that it depends only on the relative angle.
Writing ${\bf k}=k(\cos\phi,\sin\phi)$ (see Fig. \ref{f-axis}), we have thus
$k\approx k'\approx k_{\rm F}$, and we can write $|{\bf q}|=2k_{\rm
F}\sin[(\phi-\phi')/2]$ and $\alpha^2F({\bf
k-k'},\Omega)=\alpha^2F(\phi-\phi',\Omega)$. Moreover, the condition
$|\mu|\gg \omega_{\rm max},\Gamma_{\rm qp}$ allows us also to restrict
ourselves to a single electron (hole) cone for $\mu>0$ ($\mu<0$), with a
significant simplification of the calculations [see Eq.\ (\ref{gdress})
below].

Before discussing the one-particle self-energy,
let us briefly summarize the properties of the
non interacting system whose Green's function per spin and valley,
in the Matsubara space, reads:
\begin{eqnarray}
\lb{gre0}
\hat{G}^0({\bf k},i\omega_n)
&=&
\frac{1}{(i \hbar\omega_n+\mu)\hat{I}-\hbar v_{\rm F} {\bf k}
\cdot\mbox{\boldmath$\sigma$}}
\nonumber\\
&=&
\frac{(i \hbar\omega_n+\mu)\hat{I}+\hbar v_{\rm F} {\bf k}
\cdot\mbox{\boldmath$\sigma$}}
{[(i \hbar\omega_n+\mu)]^2-(\hbar v_{\rm F} k)^2}.
\end{eqnarray}
We can expand the Green's function
in the Pauli matrix basis,
\be
\lb{defg}
\hat{G}^0({\bf k},i\omega_n)
=
\sum_{i=I,x,y} G_i^0({\bf k},i\omega_n)\hat{\sigma}_i.
\ee
It is easy to see, from Eqs. (\ref{gre0})-(\ref{defg}),
that the diagonal part $G_I^0({\bf k},i\omega_n)$ depends only
on the $\e=\hbar v_{\rm F}k$ ($k=|{\bf k}|$)
while the off-diagonal components depend also
on the angle $\phi$.
In particular, we can write explicitly
\begin{eqnarray*}
G_I^0({\bf k},i\omega_n)
&=&
G_+^0(\epsilon,i\omega_n),
\\
G_x^0({\bf k},i\omega_n)
&=&
G_-^0(\epsilon,i\omega_n)\cos\phi,
\\
G_y^0({\bf k},i\omega_n)
&=&
G_-^0(\epsilon,i\omega_n)\sin\phi,
\end{eqnarray*}
where
\begin{eqnarray*}
G_\pm^0(\epsilon,i\omega_n)
&=&
\frac{1}{2}
\left[
\frac{1}{i \hbar\omega_n+\mu-\e}
\pm
\frac{1}{i \hbar\omega_n+\mu+\e}
\right].
\end{eqnarray*}

\section{One-particle self-energy}
\label{s:qp}

\begin{figure}[t]
\includegraphics[scale=0.5]{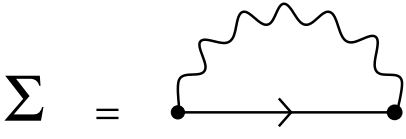}
\caption{Diagrammatic representation of the
electron-phonon self-energy. Graphical elements as
in Fig. \ref{f-kernel}.}
\label{f-self}
\end{figure}

Let us now consider the one-particle self-energy
for the electron-phonon interaction. As mentioned,
we are interested here on the limit
$|\mu|\gg \omega_{\rm max}$. In this regime Migdal's theorem\cite{migdal}
assures the validity of the standard mean-field like
Eliashberg's theory.\cite{eliashberg} The corresponding self-energy
is thus diagrammatically depicted in Fig. \ref{f-self},
and it can be written in Matsubara space as:
\begin{eqnarray}
\lb{self}
\hat{\Sigma}({\bf k},n)
&=&
T\sum_{{\bf k'},m}
W_{\bf k-k'}(n-m)
\hat{G}({\bf k'},m).
\end{eqnarray}
where $\hat{G}({\bf k},n)=\hat{G}({\bf k},i\omega_n)$,
$\hat{\Sigma}({\bf k},n)=\hat{\Sigma}({\bf k},i\omega_n)$
and $W_{\bf k-k'}(n-m)=W_{\bf k-k'}(i\omega_n-i\omega_m)$.

From Dyson's equation we can write:
\begin{eqnarray*}
\hat{G}({\bf k},n)
&=&
\frac{1}{(i \hbar\omega_n+\mu)\hat{I}-\hbar v_{\rm F} {\bf k}
\cdot\mbox{\boldmath$\sigma$}-\hat{\Sigma}({\bf k},n)}.
\end{eqnarray*}

Since Eq. (\ref{self}) is a convolution both in momentum
and frequency space,
it is easy to show
that the matrix self-energy admits the analogous decomposition (\ref{defg})
of the Green's function (\ref{defg}), namely 
$$
\hat{\Sigma}({\bf k},n)
=
\sum_{i=I,x,y} \Sigma_i({\bf k},n)\hat{\sigma}_i,
$$
where
$\Sigma_I({\bf k},n)=\Sigma_I(\e,n)$,
$\Sigma_x({\bf k},n)=\Sigma_x(\e,\phi,n)=\Sigma_{\rm off}(\e,n)\cos\phi$,
$\Sigma_y({\bf k},n)=\Sigma_y(\e,\phi,n)=\Sigma_{\rm off}(\e,n)\sin\phi$,
where the label ``off' characterizes the {\em off}-diagonal elements
of the self-energy.

Using the Dyson's equation, we can write once more
$G_I({\bf k},n)=G_+(k,n)$, $G_x({\bf k},n)=G_-(\e,n)\cos\phi$, $G_y({\bf
k},n)=G_-(\e,n)\sin\phi$, with
\begin{eqnarray}
G_\pm(\e,n)
&=&
\frac{1}{2}
\left[
\frac{1}{i \hbar\omega_n+\mu-\e
-\Sigma_I(\epsilon,n)-\Sigma_{\rm off}(\epsilon,n)}
\right.
\nonumber\\
\lb{gpm}
&&
\left.
\pm\frac{1}{i \hbar\omega_n+\mu+\e
-\Sigma_I(\epsilon,n)+\Sigma_{\rm off}(\epsilon,n)}
\right].
\end{eqnarray}

The calculation of Eq.\ \pref{self} is straightforward in the case of
intrinsically doped graphene. As discussed above, the relevant electron
momenta are here restricted on the Fermi surface, ${\bf k}\approx k_{\rm
F}$ ($\e=\mu$), so that the self-energy (\ref{self}) depends only on the
angular part of the $\bk$ vector, $\Sigma_i(\e,\phi,n)\approx
\Sigma_i(\phi,n)$.  We can also split the sum over ${\bf k'}$ in
Eq. (\ref{self}) in its energy and angular degrees of freedom, $\sum_{\bf
k}\longrightarrow \int_{-\pi}^{\pi} d\phi/2\pi \int_{0}^{W} d\epsilon
N(\epsilon)$, where $W$ is the electron bandwidth, and
$N(\epsilon)=\epsilon V/2\pi\hbar^2 v_{\rm F}^2$ is the density
of states per spin and per valley, and V is the unit-cell volume.
Thus we have
\be 
\lb{selfall}
\Sigma_i(\phi,n)=T\sum_m \int
\frac{d\phi'}{2\pi}W_{\phi-\phi'}(n-m)\chi_i(\phi') G_{i,\rm loc}(m),
\ee
where $\chi_I=1$, $\chi_x(\phi)=\cos\phi, \chi_y(\phi)=\sin\phi$, and
$G_{i,\rm loc}(m)$ is the local (${\bf k}$-averaged) Green's function:
$G_{I,\rm loc}(m)=G_{+,\rm loc}(m)=
\int N(\e) d\e G_+(\e,m)$,
$G_{x,y,\rm loc}(m)=G_{-,\rm loc}(m)=\int N(\e) d\e G_-(\e,m)$.

It is useful at this stage to introduce the
basis of the two-dimensional spherical harmonics
\bea
\psi_\a(\phi)&=&e^{i\a\phi},
\hspace{5mm} \alpha=0,\pm 1,\pm 2, \dots,
\nonumber
\eea
so that we can decompose any generic angle-dependent function $S(\phi)$
on this basis,
\bea
\lb{defcomp}
S(\phi)
&=&
\sum_\a S_\a \psi_\a(\phi),
\eea
where
\bea
S_\alpha
&=&
\int
\frac{d\theta}{2\pi}S(\phi)\psi_\alpha^*(\phi).
\eea

Using the definition (\ref{wkernel})
of the electron-phonon kernel,
we can write the diagonal and off-diagonal
components of the self-energy in Eq. (\ref{selfall}) as
\begin{eqnarray}
\lb{selfdia}
\Sigma_I(n)
&=&
T\sum_m\int d\Omega 
\frac{2\Omega \, \alpha^2F_0(\Omega)}
{\Omega^2+(\omega_n-\omega_m)^2}
G_{+,\rm loc}(m),\\
\lb{selfoff}
\Sigma_{\rm off}(n)
&=&T\sum_m\int d\Omega 
\frac{2\Omega \, \alpha^2F_1(\Omega)}
{\Omega^2+(\omega_n-\omega_m)^2}
G_{-,\rm loc}(m).
\end{eqnarray}
where $\a^2F_\a$ are the projections
of the Eliashberg function $\a^2F(\phi)$
on the spherical harmonics $\psi_\a(\phi)$, according to the decomposition
\pref{defcomp} above.

Eqs. (\ref{selfdia})-(\ref{selfoff}) can be easily analytically continued
on the real-frequency axis using standard techniques.  Through the
imaginary part of the self-energy, we can thus define a {\em diagonal} and
an {\em off-diagonal} scattering rate, $\Gamma_{I(\rm off)}=-
\lim_{\omega\rightarrow 0} \mbox{Im}\Sigma_{I(\rm off)}(\omega+i0^+)$,
where
\begin{eqnarray}
\Gamma_I
&=&
-2\pi N(\mu)\int d\Omega
\alpha^2F_0(\Omega)[n(\beta\Omega)+f(\beta\Omega)]
\nonumber\\
&&\times
\mbox{Im}
\left[G_{+,\rm loc}(\Omega+i0^+)\right],
\label{gammaI}
\end{eqnarray}
and where 
\begin{eqnarray}
\Gamma_{\rm off}
&=&
-2\pi N(\mu)\int d\Omega
\alpha^2F_1(\Omega)[n(\beta\Omega)+f(\beta\Omega)]
\nonumber\\
&&\times
\mbox{Im}
\left[G_{-,\rm loc}(\Omega+i0^+)\right].
\label{gammaoff}
\end{eqnarray}
Here $n(x)=1/[\mbox{e}^x-1]$ and $f(x)=1/[\mbox{e}^x+1]$ are 
the Bose and Fermi factors, respectively.

For a practical evaluation of the self-energy terms, we can make use once
more of the fact that, due to the low-energy phonon-mediated scattering,
the relevant electron energy are restricted to the Fermi level.  In this
case we can write $\int_{0}^{W} d\epsilon N(\epsilon)\approx N(\mu)
\int_{0}^{W} d\epsilon$ and, assuming $\mu > 0$, it is easy to see that
only the upper Dirac cone [first term in Eq. (\ref{gpm})] is relevant.  We
have in particular $G_{+,\rm loc}(\Omega+i0^+)\approx G_{-,\rm
loc}(\Omega+i0^+) =-i\pi N(\mu)/2$, so that, just as in common metals, the
real-part of the self-energy vanishes, and we have $\Gamma_{\rm I}=2K_0$,
$\Gamma_{\rm off}=2K_1$, where
\be
\lb{defka}
K_\a=\frac{\pi N(\mu)}{2}\int d\O \a^2
F_\a(\O)[n(\beta\O)+f(\beta\O)].
\ee

For the states at the Fermi energy, which involve only
the upper band ($\mu > 0$), we can define thus a total quasi-particle
scattering rate as
\be
\lb{gammaio}
\Gamma_{\rm qp}=\Gamma_{\rm I}+\Gamma_{\rm off}
=2K_0+2K_1,
\ee
which, using Eq. (\ref{gpm}), gives the dressed Green's functions:
\be
\lb{gdress}
G_+(\e,\omega)
\approx
G_-(\e,\omega)
\approx
\frac{1}{2}\frac{1}{\hbar\omega+\mu-\e+i\Gamma_{\rm qp}}.
\ee

From Eq. (\ref{gammaI})-(\ref{gammaio}) we obtain finally:
\begin{eqnarray}
\Gamma_{\rm qp}
&=&
\pi N(\mu)\sum_{i=0,1}\int d\Omega
\alpha^2F_i(\Omega)[n(\beta\Omega)+f(\beta\Omega)],\nn\\
&=&
\pi N(\mu)\int \frac{d\theta}{2\pi}
g^2_\theta (1+\cos \theta)[n(\beta\o_\theta)+f(\beta\o_\theta)],\nn\\
\lb{gammaqp}
\end{eqnarray}
where $g_\theta^2=I|{\bf q}|=I2k_F\sin(\theta/2)$ and $\o_\theta=\hbar
v_s|{\bf q}|=2\hbar v_sk_{\rm F}\sin(\theta/2)$.  Using the explicit
expression for the $K_\a$ coefficients (see Appendix \ref{a:eliash}), we
obtain standard results with a $\Gamma_{\rm qp}=(\pi^2 N(\mu)I/\hbar
\omega_{\rm max}v_s)T^2$ in the Bloch-Gr\"uneisen regime ($T \lesssim \o_{\rm
max}$) and $\Gamma_{\rm qp}=(\pi N(\mu)I/\hbar v_s)T$ in the high-temperature
$T \gtrsim \o_{\rm max}$ regime.

It is important to note that the total quasi-particle
scattering rate arises from both the diagonal and off-diagonal
contributions of the self-energy.
Indeed, the explicit angular dependence shown in the
second line of Eq. \pref{gammaqp} points out that the
off-diagonal terms, giving rise to the $\cos \theta$,
are fundamental in order to recover the usual $(1+\cos\theta)$ factor,
which accounts for the well-known absence of backscattering in graphene due to
chirality.\cite{stauber,hwang_vs,hwang_ph}

\section{DC conductivity}
\label{s:dccond}

\begin{figure}[t]
\includegraphics[scale=0.4]{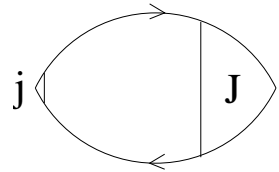}
\caption{Diagrammatic representation of the
the current-current response function $\Pi$.
The small empty triangle on the left
represents the bare current vertex $\hat{j}({\bf k})$ while
the big triangle on the right is the
renormalized current vertex function $\hat{J}({\bf k};n,n+m)$.}
\label{f-bubble}
\end{figure}

We implement now a full quantum treatment to
evaluate the DC conductivity $\sigma$ in the
presence of vertex current renormalization. 
We consider thus
the current-current response function \cite{holstein,mahan_book}
\begin{eqnarray}
\Pi(m)
&=&
\frac{N_sN_K}{V}
T\sum_{{\bf k},n}
\mbox{Tr}\left[
\hat{j}({\bf k})
\hat{G}({\bf k},n)
\right.
\nonumber\\
\lb{bubble}
&&\left.\times
\hat{J}({\bf k};n,n+m)
\hat{G}({\bf k},n+m)
\right],
\end{eqnarray}
where $N_s,N_K=2$ are respectively the spin and valley degeneracies,
$j({\bf k})=ev_{\rm F}\hat{\sigma}_x$ is the bare current operator along
the $x$-axis, and $\hat{J}({\bf k};n,n+m)$ is the fully renormalized vertex
current.  
The diagrammatic representation of the current-current response function
is shown in Fig. \ref{f-bubble}.
The DC conductivity will be obtained as the limit
$\sigma=-\lim_{\omega \rightarrow 0} \mbox{Im}\Pi(\omega+i0^+)/\omega$
after the analytical continuation of the current-current response function
$\Pi$ on the real frequency axis.

It is useful to remark here the importance
in transport properties
of the current renormalization processes
which account for the backflow
cloud associated to the current of the quasiparticles.
As discussed in the introduction,
one of the main effects of such current renormalization
in normal metals
is to give rise to the angular factor $(1-\cos\theta)$
which differentiates the transport
scattering rate $\Gamma_{\rm tr}$ from the quasi-particle
one $\Gamma_{\rm qp}$.
We can understand this result by noting that,
in normal (isotropic) metals,
the bare and 
the renormalized currents, ${\bf j}$ and ${\bf J}$ respectively,
are both proportional to
the velocity, i.e. they point in the direction of the momentum $\bk$.
Thus, one can write ${\bf J}=\Lambda {\bf k}$, where
the vertex function $\L$  can be computed for instance
in the ladder approximation. The self-consistent solution
gives thus:\cite{mahan_book}
\be
\lb{classical}
\langle \Gamma_{\rm tr}\rangle_\theta
\simeq
\left\langle \Gamma_{\rm  qp} \left(
1-\frac{\bk\cdot \bk'}{k^2}\right)\right\rangle_\theta,
\ee
where $\bk,\bk'$ are the momenta of the scattered electrons, with modulus
equal to $k_{\rm F}$ but different directions, and $\langle \dots
\rangle_\theta$ indicates the angular integration.
This leads to the usual additional
factor $(1-\bk\cdot \bk'/k^2)=(1-\cos\theta)$ in the angular average of the
transport scattering time, which reproduces by means of a full quantum
treatment the well-known semiclassical Boltzmann result.

While deriving
the result \pref{classical}, a crucial ingredient is the proportionality
between $\bJ$ and the momentum $\bk$. In graphene, where the
energy-momentum dispersion is linear, such a relation clearly does not
hold. However, as it is evident already at the level of the bare current
$\hat j$, the matrix structure plays the analogous role of the momentum
dependence in ordinary metals, so that $\hat j_x$ for example is a matrix
proportional to $\hat \sigma_x$. As far as the renormalized current is
concerned, the analogous of the momentum dependence in the ordinary metals
becomes now a decomposition of $\hat J$ in the Pauli-matrices components,
by means of dimensionless function $\hat{\Lambda}$ defined by the relation
$\hat{J}({\bf k};n,n+m)= ev_{\rm F}\hat{\Lambda}({\bf k};n,n+m)$. As we
shall see in what follows, in graphene such a matrix structure compensates
- in a non-trivial way - the lack of momentum dependence of the Fermi
velocity, and leads once more to the Boltzmann result.

Aiming on focusing on the matricial structure
of the current function,
in the following we shall make use for this quantities
of the same approximations employed for
the self-energy.  In particular, taking into account that the
electron-phonon interaction gives rise only to low-energy scattering, we
can approximate ${\bf k}\simeq k_{\rm F}$ in the vertex function and retain
only the angular dependence.  In addition, one can decompose the vertex
function in the basis of the Pauli-matrices,
\be
\lb{dec}
\hat \Lambda(\phi;n,n+m)=\sum_i \Lambda_i(\phi;n,n+m)\hat \sigma_i ,
\ee
and we can expand the $\Lambda_i(\phi)$ functions in terms of the
spherical harmonics components as in Eq.\ \pref{defcomp}:
$\Lambda_i(\phi;n,n+m)=\sum_\a \Lambda_i^\a(n,n+m)\psi_\a(\phi)$.
Inserting Eq. \pref{dec} in Eq.\ \pref{bubble}, and assuming once
more, for doped graphene,
$G_+\approx G_-$ (neglect of inter-band scattering),
we obtain thus the general structure:
\be
\lb{pi}
\Pi(m)=\frac{2e^2 v_{\rm F}^2 N_K N_S}{V} T\sum_n \L^{\rm tot}(n,n+m)
b(n,n+m),
\ee
where $b(n,n+m)= N(\mu)\int d\e G(\e,n)G(\e,n+m)$ (since $G_+=G_-$ we have
further dropped the index ``$\pm$'') and where $\L^{\rm
tot}(n,n+m)=\sum_{i,\alpha}c_i^\alpha \L_i^\alpha(n,n+m)$.  The
$c_i^\alpha$ are numerical coefficients which arise from the angular
average over $\phi$.  Since in the bubble \pref{bubble} one has the product
of two Green's functions, it is easy to realize that the angular average
will involve at most harmonics up to the second order. Indeed, an explicit
calculation shows that the only non vanishing terms are 
\bea 
\lb{eqx}
c_x^0&=& c_I^1=c_I^{-1}=1, \\
\lb{eqpar} 
c_x^2&=&c_x^{-2}=-ic_y^2=ic_y^{-2}=\frac{1}{2}.  
\eea

Once the analytical
continuation $i\omega_m\ra \omega+i0^+$ of the bubble \pref{pi} is
performed and the limit $\omega\ra 0$ is evaluated to compute the DC
conductivity, one can see\cite{mahan_book} that only the advanced-retarded
parts of the vertex function and of the $b$-function contribute, so that:
\begin{eqnarray}
\lb{sxx}
\sigma
&\simeq&
\frac{N_K N_s\hbar e^2v_{\rm F}^2}{\pi V}
b_{\rm RA}(0)\L^{\rm tot}_{\rm RA}(0),
\end{eqnarray}
where
$b_{\rm RA}(\epsilon)=
\lim_{\omega \rightarrow 0}b(\omega+i0^+,\omega-i0^+)$, and 
$\L^{\rm tot}_{\rm RA}(0)
=\lim_{\omega \rightarrow 0}\L^{\rm tot}(\omega+i0^+,\omega-i0^+)$.
From Eq.\ \pref{gdress} one easily finds
$$
b_{\rm RA}(0)=\frac{N(\mu)\pi}{4\Gamma_{\rm qp}},
$$
which gives the general expression for the DC conductivity in the presence
of vertex corrections:
\begin{eqnarray}
\lb{sdc}
\sigma
&=&
\frac{N_K N_s \hbar e^2v_{\rm F}^2 N(\mu)}{4V\G_{\rm tr}},
\end{eqnarray}
where we define the {\em transport} scattering rate
\be
\G_{\rm tr}
=
\frac{\G_{\rm qp}}{\L^{\rm tot}_{\rm RA}(0)}.
\ee

In the absence of vertex corrections ($\L^{\rm tot}_{\rm RA}(0)=1$) one
recognizes in Eq.\ \pref{sdc}, as mentioned above in the case of ordinary
metals, the standard result of the conductivity in the
bare-bubble approximation,\cite{ando_transport,sharapov} where the
scattering time for transport coincides with the quasi-particle scattering
time $\sigma=e^2 \mu/(2\pi\hbar \Gamma_{\rm qp})$.

\begin{figure}[t]
\includegraphics[scale=0.7]{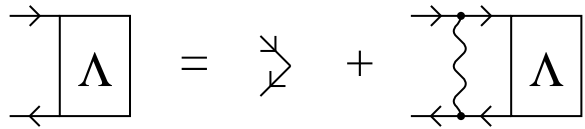}
\caption{Diagrammatic representation of the
current vertex function.
Graphical elements as
in Fig. \ref{f-kernel}.
The ladder approximation is enforced by the Migdal's 
theorem valid for
$|\mu| \gg \omega_{\rm max}$.}
\label{f-ladder}
\end{figure}

To compute Eq.\ \pref{sxx} we need then to calculate the vertex
corrections. In the doped graphene regime we are interested in,
this aim is made easy once more by the Migdal's theorem  which enforces
the validity of the mean-field theory.
The corresponding vertex function $\hat \Lambda(\bk;n,n+m)$
can be evaluated thus within
the self-consistent ladder approximation (see Fig. \ref{f-ladder}),
namely:\cite{mahan_book}
\begin{eqnarray}
&&\hat{\Lambda}({\bf k};n,n+m)
=
\hat{\sigma}_x
+
T\sum_{{\bf k'},l}
W_{\bf k-k'}(n-l)
\nonumber\\
\lb{gammaself}
&&\times
\hat{G}({\bf k'},l)
\hat{\Lambda}({\bf k'};l,l+m)
\hat{G}({\bf k'},l+m).
\end{eqnarray}

In the same limit $|\mu| \gg \omega_{\rm max}$,
we can also employ the approximations implemented 
for the self-energy, namely
$|{\bf k}|\approx |{\bf k'}|\approx k_{\rm F}$,
$\sum_{\bf k'} \longrightarrow N(\mu) \int_{-\pi}^{\pi} d\phi'/2\pi
\int_{-\infty}^\infty d\e'$, and we can also set
$\hat{\Lambda}({\bf k}'_{\rm F};l,l+m)=
\hat{\Lambda}(\phi';l,l+m)$
in the right-hand side of Eq. (\ref{gammaself}).
Expanding again the $\hat{\Lambda}(\phi';l,l+m)$
in terms of the Pauli basis and
of the spherical harmonics,
we end up with the following set of equations
\begin{eqnarray}
\Lambda_x^\a(n,n+m)
&=&
\delta_{\alpha,0}
+
T\sum_{l,j,\beta}
c_j^\beta
W_{\alpha}(n-l)
\nonumber\\
\lb{lfx}
&&\times
b(l,l+m)
\Lambda_j^{\beta+\alpha}(l,l+m),
\\
\Lambda_I^\a(n,n+m)
&=&
T\sum_{l,j,\beta}
d_j^\beta
W_{\alpha}(n-l)
\nonumber\\
&&\times
b(l,l+m)
\Lambda_j^{\beta+\alpha}(l,l+m),
\\
\Lambda_y^\a(n,n+m)
&=&
T\sum_{l,j,\beta}
f_j^\beta
W_{\alpha}(n-l)
\nonumber\\
\lb{lfy}
&&\times
b(l,l+m)
\Lambda_j^{\beta+\alpha}(l,l+m),
\end{eqnarray}
where the coefficients $c_i^\a$ are defined in Eqs. (\ref{eqx})-(\ref{eqpar}),
and the only non zero terms of the coefficients 
$d_i^\a$, $f_i^\a$ are:
\bea
d_I^0&=&2,
\nonumber \\
d_x^1&=&d_x^{-1}=-id_y^1=id_y^{-1}=1,\nonumber \\
f_y^0&=& -if_I^1=if_I^{-1}=1,\nonumber\\
-if_x^2&=&if_x^{-2}=-f_y^2=-f_y^{-2}=\frac{1}{2}.
\nonumber
\eea

We consider now the analytical continuation of Eqs. (\ref{lfx})-(\ref{lfy})
on the real axis.  We are interested on the quantity $\Lambda_{i,\rm
RA}^\a(0)$.  We can apply the usual standard procedures for the analytical
continuation\cite{mahan_book} to each element on the right side of
Eqs. (\ref{lfx})-(\ref{lfy}), and we get, for instance, 
\bea \Lambda_{x,\rm
RA}^\a(0) &=& \delta_{\alpha,0} + \frac{\pi N(\mu)}{2} \int d\Omega
\alpha^2F_\a(\Omega) \nonumber\\ &&\times [n(\beta\Omega)+f(\beta\Omega)]
\left[
\sum_{j,\beta}
c_j^\beta
\frac{\Lambda_{j,\rm RA}^{\beta+\alpha}(\Omega)}{\Gamma_{\rm qp}(\Omega)}
\right].
\label{mygamma}
\eea
Similar expressions hold true for $\Lambda_{I,\rm RA}^\a(0)$,
$\Lambda_{y,\rm RA}^\a(0)$.

The quantities $\Lambda_{j,\rm RA}^\beta(\Omega)$,
$\Gamma_{\rm qp}(\Omega)$ in the right side of Eq. (\ref{mygamma})
have a significant $\Omega$ variation over an electronic range of energies,
whereas the Eliashberg function
$\alpha^2F_\a(\Omega)$ limits the energy integration up to the phonon scale
$\omega_{\rm max}$. In this energy range we can then approximate in the
integral of Eq.\ \pref{mygamma}, 
$\Lambda_{j,\rm RA}^\beta(\Omega)\simeq
\Lambda_{j,\rm RA}^\beta(0)$, 
$\Gamma_{\rm qp}(\Omega) \simeq\Gamma_{\rm qp}(0)=\Gamma_{\rm qp}$,
and we obtain a simple set of {\em algebraic} relations for the vertex
function at zero frequency,
\begin{eqnarray}
\Lambda_{x,\rm RA}^\a(0)
&=&
\delta_{\alpha,0}
+
\frac{K_\a}{\Gamma_{\rm qp}}
\sum_{j,\beta}
c_j^\beta
\Lambda_{j,\rm RA}^{\beta+\alpha}(0),
\nonumber
\\
\Lambda_{I,\rm RA}^\a(0)
&=&
\frac{K_\a}{\Gamma_{\rm qp}}
\sum_{j,\beta}
d_j^\beta
\Lambda_{j,\rm RA}^{\beta+\alpha}(0),
\nonumber
\\
\Lambda_{y,\rm RA}^\a(0)
&=&
\frac{K_\a}{\Gamma_{\rm qp}}
\sum_{j,\beta}
f_j^\beta
\Lambda_{j,\rm RA}^{\beta+\alpha}(0).
\nonumber
\end{eqnarray}

Despite its apparent complexity, such a system of equations admits a
simple solution.  In particular, by exploiting the symmetric/antisymmetric
properties for $\a\ra -\a$ of the $x,I$ and $y$ components, respectively,
one can see that only three independent components are not zero, i.e.
$\Lambda_{x,\rm RA}^0$, $\Lambda_{x,\rm RA}^2$, $\Lambda_{I,\rm RA}^1$,
with the relations, $\Lambda_{x,\rm RA}^2 = \Lambda_{x,\rm RA}^{-2} =
i\Lambda_{y,\rm RA}^{2} = -i\Lambda_{y,\rm RA}^{-2}$, $ \Lambda_{I,\rm
RA}^1 = \Lambda_{I,\rm RA}^{-1}$.  The system can be further simplified by
noting that it can be rewritten in terms of a single self-consistent
equation for $\Lambda_{\rm RA}^{\rm tot}$:
$$
\Lambda_{\rm RA}^{\rm tot} =
1+
\frac{K_0+2K_1+K_2}{\Gamma_{\rm qp}}\Lambda_{\rm RA}^{\rm tot}, 
$$
whose solution gives
\be 
\lb{vtot}
\Lambda_{\rm RA}^{\rm tot}(0) = \frac{\Gamma_{\rm
qp}}{\Gamma_{\rm qp}-K_0-2K_1-K_2}.
\ee
Finally, by expressing the quasi-particle
scattering rate $\Gamma_{\rm qp}$ as a function of the
$K_\a$ terms as in Eq.\ \pref{gammaio} we obtain:
\be
\lb{gammatrf}
\Gamma_{\rm tr}
=
K_0-K_2.
\ee

Eqs. (\ref{vtot}), (\ref{gammatrf}) are the main result of the present paper.
Their physical insight appears clearly if we express
the transport scattering rate in terms of the microscopic
electron-phonon interaction, in analogy with the result \pref{gammaqp} for
the quasiparticle scattering rate:
\begin{eqnarray}
\G_{\rm tr}&=&
K_0-K_2\nn\\
&=&\pi N(\mu)\int \frac{d\theta}{2\pi}
g^2_\theta \frac{1-\cos 2\theta}{2}[n(\beta\o_\theta)+f(\beta\o_\theta)]\nn\\
&=&{\pi N(\mu)}\int \frac{d\theta}{2\pi}
g^2_\theta (1-\cos\theta)(1+\cos\theta)
\nonumber\\
\lb{gammatr}
&&\times
[n(\beta\o_\theta)+f(\beta\o_\theta)].
\end{eqnarray}
Taking into account the expression of the quasi-particle scattering rate
$\G_{\rm qp}$ in Eq. (\ref{gammaqp}), we can write Eq. (\ref{gammatr}) as
\be
\langle \G_{\rm tr} \rangle_\theta = \langle \G_{\rm
qp}(1-\cos \theta) \rangle_\theta.  
\ee

\begin{figure}[t]
\includegraphics[scale=0.38]{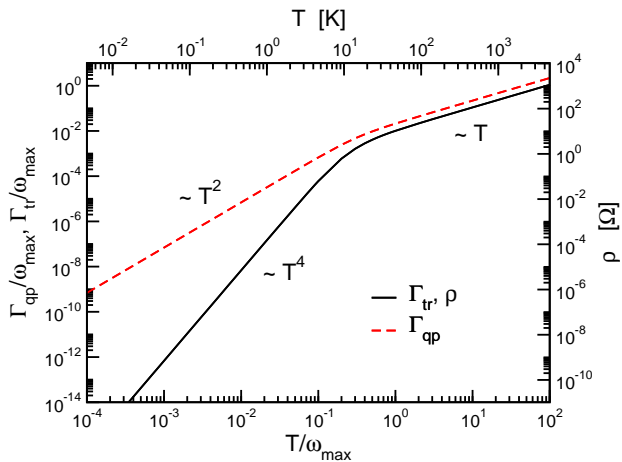}
\caption{(Color online). Temperature behavior
of the quasi-particle scattering rate
$\G_{\rm qp}$ and of the transport scattering rate $\G_{\rm tr}$
for $\mu=0.1$ eV, corresponding to $\o_{\rm max}\approx 4$ meV.
The transport scattering rate $\G_{\rm tr}$ can be also expressed as
a function of resistivity,
$\rho=4V\Gamma_{\rm tr}/N_K N_s \hbar e^2 v_{\rm F}^2N(\mu)$,
as plotted on the right-side scale.}
\label{f-rho_T}
\end{figure}

This analysis shows thus that a fully quantum derivation of the DC
conductivity yields in doped graphene the same result than the standard
Boltzmann theory.  In particular we can see that, although the vertex
function is {\bf k}-independent in graphene, the effect of current vertex
corrections is to add, just as in common systems, the additional angular
factor $1-\cos\theta$ in the phase space probed by the electron-phonon
interaction, giving rise to a suppression of the {\em forward} scattering.
The effects of such angular factor can be remarkably traced
in the temperature dependence of the 
transport scattering rate $\G_{\rm tr}=K_0-K_2$
(and hence of the resistivity $\rho=1/\sigma$)
compared with the quasi-particle one $\G_{\rm qp}=2(K_0+K_1)$.
An analytical derivation of the $K_\a$ coefficients
is reported in Appendix \ref{a:eliash} and the corresponding
temperature dependence of $\G_{\rm tr}$ and $\G_{\rm qp}$
is shown in Fig. \ref{f-rho_T}.
These results are pretty well consistent with the standard
Boltzmann theory.\cite{hwang_vs,hwang_ph}
In the high temperature limit $T \gg \o_{\rm max}$
only the $K_0$ component survives, with $K_0=\pi N(\mu)IT/2\hbar v_s$
(see Appendix \ref{a:eliash}), so that $\G_{\rm tr}, \G_{\rm qp}\sim T$.
On the other hand in the Bloch-Gr\"uneisen regime,
just as in common metals, we recover
the usual $\G_{\rm tr} \sim T^4$ while $\G_{\rm qp} \sim T^2$.
This different  dependence can be understood by noting that
the low temperature behavior of $\G_{\rm qp}$ in Eq. (\ref{gammaio})
is dominated
by the leading order $\sim T^2$ of $K_0$, $K_1$, whereas
the leading orders $K_0$, $K_2$ in Eq. (\ref{gammatrf})
cancel out so that
the temperature behavior of $\G_{\rm tr}$ in this regime
stems from the higher order $\sim T^4$ contributions.

\section{Breaking the chirality}

In the previous Section we have evaluate the DC conductivity
in chiral doped graphene
by using a quantum approach based on the Kubo formula.
We have shown that, although the vertex function is momentum
independent, the Boltzmann results are fully reproduced,
even for what concerns the presence of the so-called
angular transport factor $1-\cos\theta$ which suppresses
forward scattering.
We would like to stress however that
the resulting validity of the Boltzmann theory
is by no means trivial.
Indeed, as we have shown, the robustness of the Boltzmann results
is due to the chiral structure itself of graphene,
$\hat{H}_{\bf k}=\hbar v_{\rm F} {\bf k}
\cdot\mbox{\boldmath$\sigma$}$,
which translate in the direction
of the Pauli matrix pseudospin $\mbox{\boldmath$\sigma$}$
the role which in common metals is played
by the momentum ${\bf k}$-direction.

Such observation rises the question about the validity
of the Boltzmann result when the chiral structure
of graphene is affected by external fields.
This happens for instance in the  
presence of a weak sublattice inequivalence, where
an energy potential difference between the two sublattices,
besides  opening a gap at the
Dirac point,\cite{manes,gusynin_review2007}
gives rise also to a mixing of the chiral eigenstates and to
a loss of chirality close to the Dirac point.
Such reduced role of the chirality close to the Dirac point
has been traced, for instance,
in Ref. \onlinecite{mucha,rotenberg_njp07} in  the
intensity profile along the constant-energy contours
probed by angle-resolved photoemission.\cite{mucha,rotenberg_njp07}
The validity of the Boltzmann results even in this context
needs thus to be revised.

To this aim in this section we consider doped graphene in the presence of
a weak inequivalence between the two carbon sublattices,
parametrized in terms of a different sublattice potential $\Delta$:
\begin{eqnarray}
\lb{hamgap}
\hat{H}_{\bf k}
&=&
\hbar v_{\rm F}
{\bf k} \cdot\mbox{\boldmath$\sigma$}+\Delta\hat{\sigma}_z.
\end{eqnarray}
The energy spectrum is easily obtained from (\ref{hamgap}),
$E_k=\sqrt{(\hbar v_{\rm F}k)^2+\Delta^2}$, showing the
opening of a gap at the Dirac point.

Let us consider for the moment the case of a non-interacting system.
The presence of $\Delta$ induces an explicit $\propto \hat{\sigma}_z$
component in both the Green's function and self-energy.
Eq. (\ref{defg}) is thus generalized as
$\hat{G}({\bf k},n) =\sum_{i=I,x,y,z} G_i({\bf k},n)\hat{\sigma}_i$,
where
$G_I({\bf k},n)=G_+(E,n)$, $G_x({\bf k},n)=\gamma_{\rm off}(E)
G_-(E,n)\cos\phi$, $G_y({\bf k},n)=
\gamma_{\rm off}(E)G_-(E,n)\sin\phi$,
$G_z({\bf k},n)=\gamma_z(E)G_-(E,n)$,
with
\begin{eqnarray*}
G_\pm(E,n)
&=&
\frac{1}{2}
\left[
\frac{1}{i\hbar \omega_n+\mu-E}
\pm
\frac{1}{i\hbar \omega_n+\mu+E}
\right],
\end{eqnarray*}
$\gamma_{\rm off}(E)=\epsilon/E$,
$\gamma_z(E)=\Delta/E$.
Here we denote $\epsilon=\hbar v_{\rm F}k$ and
$E=\sqrt{\e^2+\Delta^2}$.

Since we are interested in the doped graphene,
where $|\mu|\gg \omega_{\rm max}$, and we assume $\mu > 0$,
we can as usual consider only the upper band, and
we can restrict the electronic
states on the Fermi surface, for which $E \approx \mu$,
$\e \approx \sqrt{1-\Delta^2/\mu^2}$,
and we have
\begin{eqnarray}
\lb{gupgap}
G_i(E,n)=\frac{\gamma_i}{2}g(E,n),
\end{eqnarray}
where $\gamma_I=1$,
$\gamma_x=\gamma_y=\gamma_{\rm off}=\sqrt{1-\Delta^2/\mu^2}$,
$\gamma_z=\Delta/\mu$ and where
$g(E,n)=1/(i\hbar\omega_n+\mu-E)$.

Plugging Eq. (\ref{gupgap}) in (\ref{self}), 
we can write
\begin{eqnarray*}
\Sigma_I(n)
&=&
\frac{T}{2}\sum_m
W(n-m)
g_{\rm loc}(m),
\\
\Sigma_{\rm off}(n)
&=&
\frac{\gamma_{\rm off}T}{2}\sum_m
W_1(n-m)
g_{\rm loc}(m),
\\
\Sigma_z(n)
&=&
\frac{\gamma_zT}{2}\sum_m
W(n-m)
g_{\rm loc}(m),
\end{eqnarray*}
where $g_{\rm loc}(m)=N(\mu)\int dE g(E,m)$. 
After the standard analytical continuation on the real-axis,
we can see as usual that the real part of self-energy vanishes
and the imaginary parts of the different components give
rise to the corresponding scattering rates
\begin{eqnarray}
\lb{gigap}
\Gamma_I&=&2K_0,\\
\Gamma_{\rm off}&=&
\gamma_{\rm off}
2K_1,\\
\lb{gzgap}
\Gamma_z&=&\gamma_z
2K_0.
\end{eqnarray}

Inserting this relations in the matrix expression
for the Green's function, after some careful derivation
described in Appendix \ref{a:gammagap}, we can write
a Green's function in the presence of interaction
in the form of Eq. (\ref{gdress}),
with
\begin{eqnarray*}
\Gamma_{\rm qp}
&=&
\Gamma_I
+
\gamma_{\rm off}\Gamma_{\rm off}
+
\gamma_z\Gamma_z
\nonumber\\
&=&
(1+\gamma_z^2)2K_0
+
\gamma_{\rm off}^22K_1
\nonumber\\
&=&
\lb{sqpgapp}
\left(1+\frac{\Delta^2}{\mu^2}\right)2K_0
+
\left(1-\frac{\Delta^2}{\mu^2}\right)2K_1.
\end{eqnarray*}
We can write thus
\begin{eqnarray}
\Gamma_{\rm qp}
&=&
\pi N(\mu)\int \frac{d\theta}{2\pi}
g^2_\theta[n(\beta\o_\theta)+f(\beta\o_\theta)]
\nonumber\\
\lb{gammaqpgap}
&&
\times
\left[
\left(1+\frac{\Delta^2}{\mu^2}\right)
+\left(1-\frac{\Delta^2}{\mu^2}\right)\cos \theta
\right].
\end{eqnarray}
Eq. (\ref{gammaqpgap}) shows in a direct way
the reduced effect of the chirality on the quasi-particle
scattering rate in the presence 
of a sublattice inequivalence. For instance we can see that
the back-scattering $\theta=\pi$
is now not completely suppressed but it is reduced
by a factor $\Delta^2/\mu^2$ with respect to the forward-scattering
$\theta=0$.
For $|\mu|=\Delta$, when the chemical potential is on the edge
of the gap,\cite{mucha} the $K_1$ angular dependence disappears
completely and the quasi-particle scattering is completely isotropic
with no effect of chirality.

Let us consider now the DC conductivity. 
Using the same derivation as in Sec. \ref{s:dccond},
we can still write a set of equations as 
(\ref{lfx})-(\ref{lfy}) with the same $c_i^\a$,
$d_i^\a$,$f_i^\a$ and rescaled quantities
$K_0\rightarrow \gamma_{\rm off}^2K_0$,
$K_1\rightarrow (1+\gamma_z^2)K_1$,
$K_2\rightarrow \gamma_{\rm off}^2K_2$,
The presence of $\propto \hat{\sigma}_z$ terms in the Green's function
gives rise in addition to a finite $z$-component $\Lambda_z^\a$.
After lengthly but straightforward calculations we can still write
the same self-consistent equation for the total vertex function
as before, with the rescaling of the $K_\a$ coefficients discussed above.
The final result is
$$
\Lambda_{\rm RA}^{\rm tot}(0)
=
\frac{\gamma_{\rm off}^2\Gamma_{\rm qp}} {\Gamma_{\rm qp}-
\gamma_{\rm off}^2K_0-2(1+\gamma_z^2)K_1-\gamma_{\rm off}^2K_2},
$$
so that
\bea
\Gamma_{\rm tr}
&=&
\frac{\Gamma_{\rm qp}-\gamma_{\rm off}^2K_0-2(1+\gamma_z^2)K_1
-\gamma_{\rm off}^2K_2}{\gamma_{\rm off}^2}
\nonumber\\
\lb{guff}
&=&
\frac{\mu^2+3\Delta^2}{\mu^2-\Delta^2}K_0
-\frac{4\Delta^2}{\mu^2-\Delta^2}K_1
-K_2.
\eea
Eq. (\ref{guff}) looks at a first sight quite dim.  However, taking into
account the explicit angular dependence of the Eliashberg functions
$\alpha^2F_\a$ and of the coefficients $K_\a$, we can re-arrange
Eq. (\ref{guff}) in the form:
\bea
\lb{scalinggap}
\langle \G_{\rm tr} \rangle_\theta
=
\left(1-\frac{\Delta^2}{\mu^2}\right)^{-1}
\langle \G_{\rm qp}(1-\cos \theta) \rangle_\theta,
\eea
where the presence of the angular term $(1-\cos \theta)$ suggests again the
validity of the Boltzmann result even in this case where the chirality of
graphene close to the Dirac point is affected by the sublattice
inequivalence. As far as the prefactor is concerned, one could in principle
justifies this as well within the Boltzmann framework by taking into
account that the the opening of the gap affects also the Fermi velocity of
the eigenstates of the Hamiltonian \pref{hamgap}. Thus, following the
Boltzmann prescription, the velocity in the basis of the eigenstates is
$v_k=\hbar^{-1}d\sqrt{(\hbar v_{\rm F}k)^2+\Delta^2}/dk =v_{\rm
F}\e_k/\sqrt{\e_k^2+\Delta^2}\approx v_F\sqrt{1-\Delta^2/\mu^2}$, where
$\e_k=\hbar v_{\rm F}k$ and $\sqrt{\e_k^2+\Delta^2}\approx \mu$ at the
Fermi level. Thus, the additional $(1-\D^2/\mu^2)$ factor, that we obtained
in Eq.\ \pref{scalinggap} associated with vertex corrections, will appear
in the Boltzmann expression for the conductivity $\sigma_{\rm B}=e^2
N(\mu)v^2/2\langle\Gamma_{B}\rangle$ associated to the $v^2$ term instead
of the scattering rate, which is simply $\langle \Gamma_{\rm
B}\rangle_\theta=\langle \G_{\rm qp}(1-\cos \theta) \rangle_\theta$.

\section{Discussion}

\begin{figure}[t]
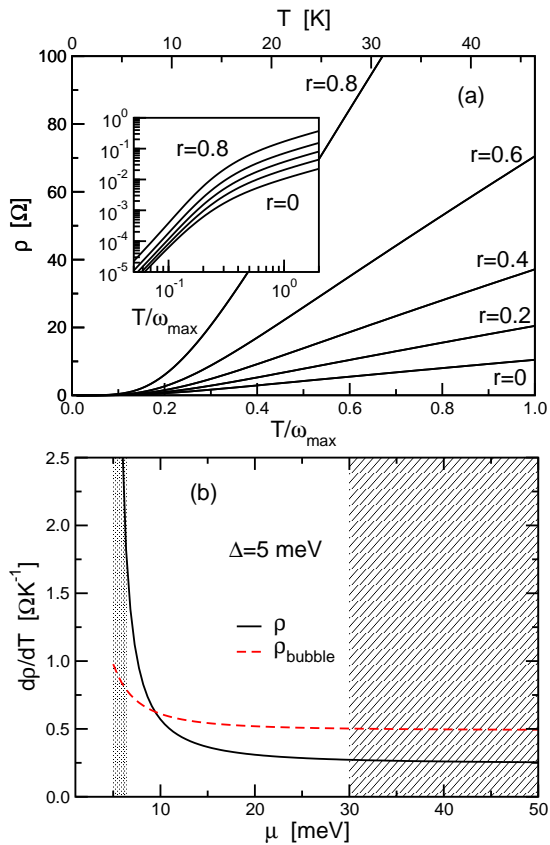

\includegraphics[scale=0.38,clip=]{f-rhogap_T.eps}
\includegraphics[angle=0,scale=0.38,clip=]{f-slope_mu.eps}
\caption{(color online): (a) Temperature dependence of the resistivity 
for different doping in the presence of
a weak sublattice inequivalence, giving rise to the gap $\Delta$.
Here $r=\Delta^2/\mu^2$.
Resistivity has been computed using the
transport scattering time in Eq. \pref{guff}. Inset: same
quantities on a log-log scale.
(b) Doping dependence of the high-temperature slope of
the resistivity in graphene in the presence of a gap
($\Delta=5$ meV).  The 
dashed (red) line is the result \pref{rhobare}
obtained with the bare bubble, 
while the solid (black) line represents the resistivity \pref{rho} computed
with full inclusion of vertex corrections. The shaded area at $\mu>30$ meV
represents the range of doping investigated in
Ref.\ \onlinecite{stormer}, while in the grey area around 5 meV
the validity of Eq. \pref{rho} fails because
$|\mu|-\Delta \le \Gamma_{\rm qp}$ ($\Gamma_{\rm qp}=1.4$ meV
has been computed here for $T=300$ K).} 
\label{f-rhogap}
\end{figure}

In the previous section we have shown that, in the limit where the chemical
potential is the largest energy scale of the system, the Boltzmann theory
is still valid for graphene even in the
presence of a weak sublattice inequivalence which gives rise
to a gap $\Delta$.
A possible outcome of the extension of
these calculations beyond this ``Boltzmann regime'' is suggested 
by the observation of the (weak) doping dependence of the
resistivity $\rho$ that follows from Eq.\ \pref{scalinggap}. As it is shown in 
Fig.\ \ref{f-rhogap}a,
both the low temperature ($T \ll \o_{\rm max}$)
and the high temperature ($T \gg \o_{\rm max}$) regimes acquire
a doping dependence, here parametrized in terms of the ratio
$r=(\Delta/\mu)^2$.
Particularly interesting appears the 
high-temperature regime where,
using Eq.\ \pref{guff} and \pref{kht}, one finds:
\bea
\lb{rho}
\rho
&=&
\frac{\mu^2+3\Delta^2}{\mu^2-\Delta^2}
\frac{\pi D^2}{4\hbar e^2 v_{\rm F}^2v_s^2\rho_m}T,
\eea
which implies a {\em significant} increase
of the linear slope as one approaches the
Dirac point $\mu \ra \D$, as shown in Fig.\ \ref{f-rhogap}b.
Observe that, if
one would use instead the single-bubble result for the DC conductivity, the
resistivity
would be proportional to the quasiparticle scattering rate $\Gamma_{\rm qp}$,
so that according to Eq.\ \pref{sqpgapp} one would obtain:
\be
\lb{rhobare}
\rho_{\rm bare}=
\left(1+\frac{\Delta^2}{\mu^2}\right)
\frac{\pi D^2}{4\hbar e^2 v_{\rm F}^2v_s^2\rho_m}T,
\ee
which has a much weaker doping dependence, as shown in Fig.\
\ref{f-rhogap}b. Remarkably, an increase of the linear slope as $\mu$
decreases has been observed in recent measurements of resistivity in doped
graphene samples.\cite{fuhrer,stormer} Indeed, while the measured
crossover from a power-law to a linear $T$-behavior points toward the
electron-phonon scattering as the source of the temperature dependence, the
fact that the slope of the linear term decreases as doping increases, and
eventually saturates at enough large doping, is unexplained within existing
theories.  Eq.\ \pref{rho} could suggest a possible mechanism for this
doping dependence, even though in the regime of validity of this equation
the quantitative variations of the slope are expected to be much smaller
than what experimentally measured. Indeed, recent tunneling experiments
performed in a similar suspended graphene sample have found a gap
at the Dirac point $\Delta\simeq 5$ meV,\cite{andrei}.
This small value of the gap, compared to the relevant values
$\mu\simeq 30-70$ meV of the chemical potential in the samples investigated in
Ref.\ \onlinecite{stormer}, would lead, in Eq.\ \pref{rho},
to a variation of the slope of only few percents,
as shown in Fig. \ref{f-rhogap}b.
However, one should keep in mind
that Eq.\ \pref{rho} has been derived in the limit
$\Gamma_{\rm qp}\ll \mu$.
Thus, only the full solution in the crossover region where the Dirac
point is approached can discriminate if the doping dependence reported in
Refs.\ \onlinecite{fuhrer,stormer} is an effect of a small gap opening
which goes beyond the Boltzmann approach. Analogously, it can
be worth also to investigate alternative scenarios\cite{bc} for the gap
opening that have been suggested by photoemission results in
epitaxial graphene,\cite{lanzara_natmat07,zhou_condmat08}
where the gap $\Delta\simeq 100$ meV is
comparable to the chemical potential in the interesting doping regime.

\section{Conclusions}

In summary, in this paper we outlined the issue of the calculation of the
current vertex corrections for electron-phonon scattering in graphene
within the Kubo formalism. While previous works investigated the role of
vertex corrections in the presence of impurity
scattering,\cite{ando_transport,falkoprl,falkoprb}
here we address for the first time
the issue of vertex corrections in graphene when the scattering mechanism
arises from electron-phonon interactions, which have a non-trivial momentum
dependence. We analyzed in particular the case of doped graphene, when
$\mu$ is the largest energy scale of the system (i.e. $|\mu|\gg\omega_{\rm
max}, \Gamma_{\rm qp}$). In this regime the calculations can be performed
explicitly, leading to an analytical derivation of the transport scattering
rate which appears in the DC conductivity. Remarkably, we found that
despite the lack of direction-dependence of the quasiparticle velocity in
graphene, the matrix structure associated to the current vertex plays a
similar role than the momentum dependence of the renormalized vertex in
ordinary metals, leading to a confirmation of the Boltzmann approach.  Such
a result is confirmed also in the presence of a sublattice inequivalence,
which leads to a gapped Dirac spectrum and introduces a doping dependence
of the resistivity that can be a promising candidate to explain existing
experimental data in doped graphene samples.  With respect to the formalism
based on the quantum kinetic equations,\cite{auslender,trushin1,trushin2}
which is also aimed to investigate the applicability of the Boltzmann
result, the present approach based on the calculation of vertex corrections
for the conductivity bubble has the main advantage of being suitable of a
direct extension at finite frequency. Such an extension will lead to a full
quantum treatment of the conductivity in graphene, and will be the subject
of a future work.

\begin{acknowledgments}
We acknowledge financial support from MIUR under the Research Program
MIUR-PRIN 2007.
\end{acknowledgments}

\appendix
\section{Harmonic components of the Eliashberg functions
and $K_\a$ coefficients}
\label{a:eliash}

In this Appendix we provide an explicit expression for the
harmonic components of Eliashberg functions $\alpha^2F_\a(\Omega)$,
and we calculate the temperature dependence of the
corresponding functions $K_\alpha(T)$.
We recall the definition of the momentum-dependent
Eliashberg function,
$$
\alpha^2F({\bf k-k'},\Omega)
=
|g_{\bf k-k'}|^2 \delta(\Omega-\omega_{\bf k-k'}),
$$
where $\omega_{\bf q} \approx \hbar v_{\rm s} |{\bf q}|$.
Following Ref. \onlinecite{hwang_ph}, we notice that
$|g_{\bf q}|^2 \propto |{\bf q}|$, so that we can write
$|g_{\bf q}|^2 = I |{\bf q}|$, where
$I=\hbar D^2/2V\rho_mv_s$, $D$ is the deformation potential,
$V$ is the volume of the two-dimensional unit cell,
and $\rho_m$ is the graphene mass density.
Typical values are $D\simeq 19$ eV, $\rho_m=7.6\times 10^{-8}$ gr/cm$^2$,
$v_s=2\times 10^6$ cm/s, $v_{\rm F}=10^6$ m/s.
Restricting the electron momenta on the Fermi surface,
$|{\bf k}|\approx |{\bf k'}|\approx k_{\rm F}$,
we can write $|{\bf q}|=2k_{\rm F}\sin[(\phi-\phi')/2]$,
and the Eliashberg function will result to depend
only on the exchanged angle $\theta=\phi-\phi'$:
\be
\alpha^2F(\theta,\Omega)
=
2Ik_{\rm F}
\sin(\theta/2)
\delta[\Omega-\omega_{\rm max} \sin(\theta/2)],
\label{ae}
\ee
where we remind
$\omega_{\rm max}=2\hbar v_s k_{\rm F}$.

The angular components $\alpha^2F_\a$ will be evaluated simply
projecting the Eliashberg function (\ref{ae}) on the
spherical harmonics $\psi_\a(\phi)=\mbox{e}^{i\a \phi}$.
We have thus:
\begin{eqnarray*}
\alpha^2F_\alpha(\Omega)
&=&
2I k_{\rm F}
\int_{0}^{2\pi}\frac{d\theta}{2\pi}
\sin(\theta/2)
\mbox{e}^{i\alpha \theta}
\delta[\Omega-\omega_{\rm max} \sin(\theta/2)]
\nonumber\\
&=&
2I k_{\rm F}
\int_{0}^{\pi}\frac{d\theta}{\pi}
\sin\theta
\mbox{e}^{i2\alpha \theta}
\delta[\Omega-\omega_{\rm max} \sin\theta].
\end{eqnarray*}

The $\delta$-function has two solutions for $\theta \in [0:\pi]$,
one for $\theta=y_\Omega$ and one for $\theta=\pi-y_\Omega$,
where $y_\Omega=\arcsin(\Omega/\omega_{\rm max})$.
We can write thus:
\begin{eqnarray}
\alpha^2F_\alpha(\Omega)
&=&
2I k_{\rm F}
\int_0^{\pi}\frac{d\theta}{\pi}
\sin\theta
\mbox{e}^{i2\alpha \theta}
\frac{1}{|\omega_{\rm max} \cos(\theta)|}
\nonumber\\
&&\times
\left[
\delta(\theta-y_\Omega)+\delta(\theta-\pi+y_\Omega)
\right]
\nonumber\\
&=&
\frac{4I k_{\rm F}}{\pi\omega_{\rm max}^2}
\frac{\Omega\theta(\omega_{\rm max}-\Omega)}
{\sqrt{1-\left(
\frac{\displaystyle \Omega}{\displaystyle \omega_{\rm max}}\right)^2}}
\nonumber\\
&&\times
\frac{1}{2}\left[
\mbox{e}^{i2\alpha y_\Omega}+\mbox{e}^{i2\alpha(\pi- y_\Omega)}
\right]
\nonumber\\
\lb{alphaa}
&=&
\frac{2I}{\pi\hbar v_s}
\frac{\O\cos(2\alpha y_\Omega) \theta(\omega_{\rm max}-\Omega)}
{\sqrt{\o_{\rm max}^2-\O^2}},
\end{eqnarray}
where
we made use of 
$\sin(y_\Omega)=\sin(\pi-y_\Omega)=\Omega/\omega_{\rm max}$,
$|\cos(y_\Omega)|=|\cos(\pi-y_\Omega)|
=\sqrt{1-(\Omega/\omega_{\rm max})^2}$. 

Inserting Eq.\ \pref{alphaa} into
the Eq.\ \pref{defka} which defines the $K_\a$ function, we can obtain their
temperature dependence. By rescaling the integration variable to $y=\b
\O$ we have:
\bea 
K_\a&=&\frac{N(\mu)I}{\hbar v_s}\frac{T^2}{\o_{\rm max}}
\int_0^{\beta\omega_{\rm max}} dy \frac{y}{\sqrt{1-(y/\beta\omega_{\rm
max})^2}}\nn\\
\lb{eqka}
&\times &\cos\left(2\alpha\arcsin\frac{y}{\beta\omega_{\rm max}}\right)
\left[n(y)+f(y)\right].
\eea
In the low-temperature limit $T \ll\o_{\rm max}$
we can retain the leading terms in powers of $y/\beta\o_{\rm max}$ in
the integrand and send the upper limit of integration to infinity,
$\beta\omega_{\rm max} \ra \infty$.
We obtain thus:
$$
K_\alpha=\frac{N(\mu)I}{\hbar v_s}\frac{T^2}{\o_{\rm max}}
\left[a_1+a_3\frac{1-4\a^2}{2}\frac{T^2}{\o^2_{\rm max}}\right],
$$
where $a_n=\int_0^\infty dy y^n [n(y)+f(y)]$, and
in particular, $a_1=\pi^2/4$ and
$a_3=\pi^4/8$.
In the opposite high-temperature limit $T \gg\o_{\rm max}$ we can
instead compute the integrand function in Eq.\ \pref{eqka} as $y\ra 0$. In
this case, due to the rapid oscillations of the cosine term, all the
harmonics vanish except the $\a=0$ one, which displays a 
linear $T$ dependence: 
\be
\lb{kht}
K_0(T)=\frac{\pi N(\mu)I}{2\hbar v_s}T.
\ee

\section{Total quasi-particle scattering rate
in the presence of sublattice inequivalence}
\label{a:gammagap}

In this Appendix we derive the effective total quasi-particle
scattering rate in graphene in the presence of sublattice inequivalence.

Let us start from the scattering rates
in Eqs. (\ref{gigap})-(\ref{gzgap}), which we can write
in the matricial form:
\bea
\hat{\Gamma}
=
\Gamma_I\hat{I}
+\Gamma_{\rm off}[\cos\phi\hat{\sigma}_x+\sin\phi\hat{\sigma}_y]
+\Gamma_z\hat{\sigma}_z.
\eea
Considering the Hamiltonian in Eq. (\ref{hamgap}),
we have then
\begin{eqnarray*}
\hat{G}^{-1}(\e,\omega)
&=&
(\hbar \omega+i\Gamma_I)\hat{I}
+(\Delta-i\Gamma_z)\hat{\sigma}_z
\nonumber\\
&&
-(\e-i\Gamma_{\rm off})[\cos\phi\hat{\sigma}_x+\sin\phi\hat{\sigma}_y].
\end{eqnarray*}
The spectral function, whose width determines the total effective
quasi-particle scattering rate, is associated with the $G_I$ term
which results:
\bea
\lb{ggigap}
G_I
(\e,\omega)
&=&
\frac{\hbar\omega+\mu+i\Gamma_I}
{(\hbar \omega+\mu+i\Gamma_I)^2
-(\e-i\Gamma_{\rm off})^2
-(\Delta-i\Gamma_z)^2}.
\eea

In order to identify the total effective
quasi-particle scattering rate,
we expand now the denominator and write, in particular,:
\begin{eqnarray*}
&&(\e-i\Gamma_{\rm off})^2
+(\Delta-i\Gamma_z)^2
\nonumber\\
&=&
\e^2+\Delta^2-\Gamma_{\rm off}^2-\Gamma_z^2
-2i(\e\Gamma_{\rm off}+\Delta\Gamma_z).
\end{eqnarray*}
Since $\e^2+\Delta^2\approx \mu$ and $\Gamma_{\rm off},
\Gamma_z \ll |\mu|$, we can neglect $\Gamma_{\rm off}^2-\Gamma_z^2$
with respect to $E^2=\e^2+\Delta^2$ and, at the same order, we can write:
\begin{eqnarray*}
&&E^2
-2i(\e\Gamma_{\rm off}+\Delta\Gamma_z)
\approx
\left[
E
-i\left(
\frac{\e}{E}\Gamma_{\rm off}
+
\frac{\Delta}{E}\Gamma_z
\right)
\right]^2
\nonumber\\
&\approx&
\left[
E
-i\left(
\sqrt{1-\frac{\Delta^2}{\mu^2}}\Gamma_{\rm off}
+
\frac{\Delta}{\mu}\Gamma_z
\right)
\right]^2.
\end{eqnarray*}

Plugging this result in (\ref{ggigap}), we can
now split the resulting Green's function
in the usual two contributions from the upper and lower band,
$$
G_I(E,\omega)
=
\frac{1}{2}
\sum_{s=\pm}
\frac{1}{\hbar\omega+\mu+i\Gamma_I
\mp E\pm i\Gamma_\pm},
$$
with
$$
\Gamma_\pm
=
\Gamma_I
\pm
\left(\sqrt{1-\frac{\Delta^2}{\mu^2}}\Gamma_{\rm off}
+\frac{\Delta}{\mu}\Gamma_z\right),
$$
so that at the Fermi level in the upper band we obtain
$$
\Gamma_{\rm qp}
=
\Gamma_I
+
\sqrt{1-\frac{\Delta^2}{\mu^2}}\Gamma_{\rm off}
+\frac{\Delta}{\mu}\Gamma_z.
$$


\begin{thebibliography}{99}


\bibitem{guinea_review}
F. Guinea, N.M.R. Peres, K.S. Novoselov, and A.K. Geim,
arXiv:0709.1163  [cond-mat] (2007).

\bibitem{ando_back}
T. Ando and T. Nakanishi,
J. Phys. Soc. Jpn {\bf 67}, 1704 (1998).

\bibitem{lee}
P.A.~Lee, \prl {\bf 71}, 1887 (1993)

\bibitem{hirschfeld}
P.J.~Hirschfeld, W.O.~Putikka, and D.J.~Scalapino, \prb  {\bf 50} 10250
(1994). 

\bibitem{homes}
Y.S. Lee, K. Segawa, Z.Q. Li, W.J. Padilla, M. Dumm,
S.V. Dordevic, C.C. Homes, Y. Ando, and D.N. Basov,  
\prb  {\bf 72} 054529 (2005).

\bibitem{mccann}
E. McCann and V.I. Fal'ko,
Phys. Rev. Lett. {\bf 96}, 086805 (2006).

\bibitem{holstein}
T. Holstein,
Ann. Phys. {\bf 29}, 410 (1964).

\bibitem{mahan_book}
G.D. Mahan,
{\em Many-Particle Physics} (Plenum, New York, 1981).

\bibitem{nomura}
K. Nomura and A.H. MacDonald,
Phys. Rev. Lett. {\bf 96}, 256602 (2006).

\bibitem{ando_screen}
T. Ando,
J. Phys. Soc. Jpn {\bf 75}, 074716 (2006).

\bibitem{hwangprl}
E.H. Hwang, S. Adam, and S. Das Sarma,
Phys. Rev. Lett. {\bf 98}, 186806 (2007).

\bibitem{adam_pnas}
S. Adam, E.H. Hwang, V. Galitski, and S. Das Sarma,
Proc. Natl. Acad. Sci. USA {\bf 104}, 18392 (2007).

\bibitem{novikov}
D.S. Novikov,
Phys. Rev. B {\bf 76}, 245435 (2007).

\bibitem{stauber}
T. Stauber, N.M.R. Peres, and F. Guinea,
Phys. Rev. B {\bf 76}, 205423 (2007).

\bibitem{peres}
N.M.R. Peres, J.M.B. Lopes dos Santos, T. Stauber,
Phys. Rev. B {\bf 76}, 073412 (2007).

\bibitem{adam_bilayer}
S. Adam and S. Das Sarma,
Phys. Rev. B {\bf 77}, 115436 (2008).

\bibitem{adam_ssc}
S. Adam and S. Das Sarma,
Solid State Communications {\bf 146}, 356 (2008).

\bibitem{hwang_vs}
E.H. Hwang and S. Das Sarma,
Phys. Rev. B {\bf 77}, 195412 (2008).

\bibitem{hwang_ph}
E.H. Hwang and S. Das Sarma,
Phys. Rev. B {\bf 77}, 115449 (2008).

\bibitem{auslender}
M. Auslender and M.I. Katsnelson,
Phys. Rev. B {\bf 76}, 235425 (2007).

\bibitem{trushin1}
M. Trushin and J. Schliemann,
Phys. Rev. Lett. {\bf 99}, 216602 (2007).

\bibitem{trushin2}
M. Trushin and J. Schliemann,
Europhys. Lett. {\bf 83}, 17001 (2008).

\bibitem{ziman}
J.M. Ziman,
G.D. Mahan,
{\em Electrons and Phonons} (Oxford University Press, New York, 1960).

\bibitem{ando_transport}
N.H. Shon and T. Ando,
J. Phys. Soc. Jpn {\bf 67}, 2421 (1998).

\bibitem{falkoprl}
E.~McCann, K.~Kechedzhi, V.I.~Fal'ko, H.~Suzuura, T.~Ando, and 
B.L.~Altshuler, Phys. Rev. Lett. {\bf 97}, 146805 (2006).

\bibitem{falkoprb}
K.~Kechedzhi, O.~Kashuba, and V.I.~Fal'ko, 
Phys. Rev. B {\bf 77}, 193403 (2008).

\bibitem{fuhrer}
J.H. Chen, C. Jang, S. Xiao, M. Ishigami, and M.S. Fuhrer,
Nat. Nanot. {\bf 3}, 206 (2008)

\bibitem{stormer}
K.I. Bolotin, K.J. Sikes, J. Hone, H.L. Stormer, and P. Kim, 
Phys. Rev. Lett. {\bf 101}, 096802 (2008).

\bibitem{mucha}
M. Mucha-Kruczy\'nski, O. Tsyplyatyev, A. Grishin, E. McCann,
V.I. Fal'ko, A. Bostwick, and E. Rotenberg,
Phys. Rev. B {\bf 77}, 195403 (2008). 

\bibitem{rotenberg_njp07}
A. Bostwick, T. Ohta, J.L. McChesney, K.V. Emtsev, T. Seyller,
K. Horn, and E. Rotenberg, 
New. Jour. of Phys. {\bf 9}, 385 (2007).

\bibitem{basov}
Z.Q. Li, E.A. Henriksen, Z. Jiang, Z. Hao, M.C. Martin, P. Kim,
H.L. Stormer, and D.N. Basov, Nat. Phys. {\bf 4}, 532 (2008).


\bibitem{castroneto}
A.H. Castro Neto and F. Guinea,
Phys. Rev. B {\bf 75}, 045404 (2007).

\bibitem{dresselhaus}
Ge.G. Samsonidze, E.B. Barros, R. Saito, J. Jiang,
G. Dresselhaus, and M.S. Dresselhaus,
Phys. Rev. B {\bf 75}, 155420 (2007).

\bibitem{manes}
J.L.~Ma{\~n}es, F.~Guinea, and M.A.H.~Vozmediano,
Phys. Rev. B {\bf 75}, 155424 (2007). 

\bibitem{park}
C.-H. Park, F. Giustino, M.L. Cohen, and S.G. Louie,
Phys. Rev. Lett. {\bf 99}, 086804 (2007).

\bibitem{bonini}
N. Bonini, M. Lazzeri, N. Marzari, and F. Mauri,
Phys. Rev. Lett. {\bf 99}, 176802 (2007).

\bibitem{calandra}
M. Calandra and F. Mauri,
Phys. Rev. B {\bf 76}, 205411 (2007). 

\bibitem{tse}
W.-K. Tse and S. Das Sarma,
Phys. Rev. Lett. {\bf 99}, 236802 (2007).

\bibitem{basko}
D.M. Basko and I.L. Aleiner,
Phys. Rev. B {\bf 77}, 041409 (2008). 

\bibitem{migdal}
A.B. Migdal, Zh. Eksp. Teor. Fiz. {\bf 34}, 1438 (1958)
[Sov. Phys. JETP {\bf 7}, 996 (1958)].

\bibitem{eliashberg}
G.M. Eliashberg, Zh. Eksp. Teor {\bf 38}, 966 (1960)
[Sov. Phys. JETP {\bf 11}, 696 (1958)].

\bibitem{sharapov}
V.P.~Gusynin, S.G.~Sharapov, and J.P.~Carbotte, 
Phys. Rev. Lett. {\bf 96}, 256802 (2006).

\bibitem{gusynin_review2007} 
V.P.~Gusynin, S.G.~Sharapov, and J.P.~Carbotte,
Int. J. Mod. Phys. B {\bf 21}, 4611 (2007).

\bibitem{andrei}
G.~Li, A.~Luican, and E.Y.~Andrei, arXiv:0803.4016.

\bibitem{bc}
L. Benfatto and E. Cappelluti,
Phys. Rev. B {\bf 78}, 115434 (2008).


\bibitem{lanzara_natmat07}
S.~Y.~Zhou,
G.~-H.~Gweon, A.~V.~Fedorov, P.~N.~First, W.~A.~de~Heer,
D.~-H.~Lee, F.~Guinea, A.~H.~Castro~Neto, and A.~Lanzara,
Nat. Mat. {\bf 6}, 770 (2007).

\bibitem{zhou_condmat08}
S.Y.~Zhou, D.A.~Siegel, A.V.~Fedorov, and A.~Lanzara,
Physica E {\bf 4}, 2642 (2008).





\end{thebibliography}
\end{document}